\documentclass[useAMS,usenatbib]{mnras/mn2e}
\usepackage{natbib}
\bibpunct{(}{)}{;}{a}{}{,}
\usepackage{graphicx}

\begin{document}

\title[A search for radio pulsars around low-mass white dwarfs]{A
  search for radio pulsars around low-mass white dwarfs}

\author[Van Leeuwen et al.]{Joeri van Leeuwen\thanks{e-mail: \it joeri@astro.ubc.ca}, 
Robert D. Ferdman, Sol Meyer and Ingrid Stairs\\
Department of Physics \& Astronomy, University of
British Columbia, 6224 Agricultural Road, Vancouver B.C. V6T 1Z1,
Canada.}

\date{Accepted Received}

\maketitle

\begin{abstract}
Low-mass white dwarfs can either be produced in low-mass X-ray
binaries by stable mass transfer to a neutron star, or in a
common-envelope phase with a heavier white dwarf companion. We have
searched 8 low-mass white dwarf candidates recently identified in the
Sloan Digital Sky Survey for radio pulsations from pulsar companions,
using the Green Bank Telescope at 340MHz. We have found no pulsations
down to flux densities of 0.6--0.8 mJy kpc$^{-2}$ and conclude that a
given low-mass helium-core white dwarf has a probability of $<
0.18\pm0.05$ of being in a binary with a radio pulsar.
\end{abstract}

\begin{keywords}
pulsars: general --- white dwarfs
\end{keywords}

\section{Introduction}\label{sec:intro}
High-precision timing of millisecond and binary pulsars has
applications ranging from setting limits on the stochastic
gravitational-wave background \citep[e.g.][]{jb03}, to studying binary
evolution, neutron-star masses and equations of state
\citep[e.g.][]{lp01}, and testing the predictions of general
relativity \citep[e.g.][]{tw89, lbk+04}.  Wide-area systematic
searches have historically been very successful in finding millisecond
pulsars \citep[e.g.][]{llb+96}. In contrast to these wide-area
surveys, directed pulsar searches use typical associations of neutron
stars with other objects to increase the chance of finding a pulsar in
a given telescope pointing. This allows direct searches to use fewer
pointings, with longer integration times and hence better
sensitivity. Directed searches towards certain object types also
elucidate their evolutionary relation with radio pulsars 
through the incidence of the associations \citep[cf.][]{sr68}.
The classic example is searching for young pulsars toward
pulsar wind nebulae and supernova remnants, often with known
central X-ray point sources \citep[e.g.][]{csl+02}, and the success of
these searches has enabled meaningful statistics on the interactions
between the radio pulsars and the supernova remnants \citep[e.g.][]{che05}.
Directed searches for recycled {\it
millisecond} pulsars have so-far been limited to globular clusters
\citep[e.g.][]{rhs+05} and a few isolated steep-spectrum targets
\citep[e.g.][]{bkh+82,nbf+95}. 

\begin{figure}
\vspace{3mm} \includegraphics[]{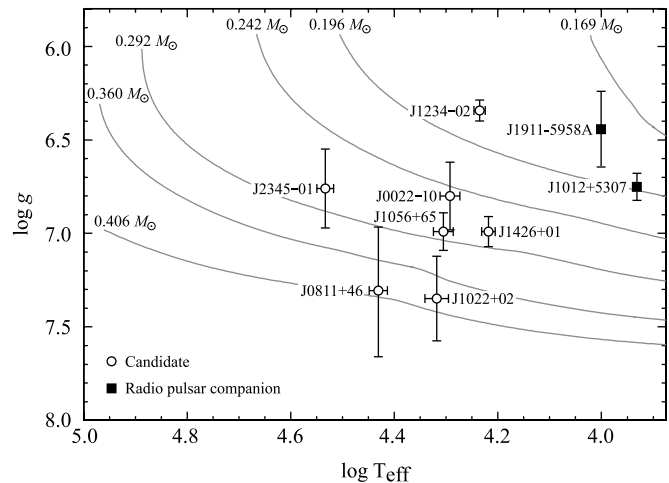}
\caption{ 
  \label{fig:LOGG_TEFF} 
  Measured temperatures and surface gravities plotted onto simulated
  evolutionary tracks for  low-mass helium-core white dwarfs of
  different mass. Shown are the 7 candidates (open circles) and 2 pulsar
  companions (filled squares) that fall within the simulated range.
  Data taken from \citet{khe+04short}, \citet{elh+06} and \citet{bvk+06}, plotted over evolutionary
  tracks from \citet{asb01} and 
  \citet{sar01} as cited in \citet{lbe+04}.
 }
\end{figure}
\vspace{10mm}

The situation has changed with the publication of catalogs of white
dwarfs from deep surveys.  In particular, \citet{khe+04short} list over
2500 spectroscopically identified white dwarfs from the first data
release of the Sloan Digital Sky Survey (SDSS, \citealt{yaa+00short}).  Among these
are a handful which appear to have very low surface gravities ($\log g
\leq 7.5$) and fairly high effective temperatures of 2--3$\times 10^4\,$K,
indicating probable masses of less than 0.45\,$M_{\odot}$ and a helium
composition (Figure \ref{fig:LOGG_TEFF}).  These objects are described
in more detail in \citet{lbe+04}, in which one object,
SDSS~J123410.37$-$022802.9, is shown to have an excellently determined
spectrum and a likely mass of $\sim 0.18$--$0.19$\,$M_{\odot}$. Such a
low-mass, helium-core white dwarf is the prototypical companion of a
highly recycled millisecond pulsar: in the low-mass X-ray binary phase
much of the white-dwarf envelope is transferred to the neutron star,
spinning it up. While this is not the only plausible explanation for
the existence of this star (a white-dwarf companion is also a
potential explanation), the match with the masses of known pulsar
companions such as those of PSRs~J1012+5307
\citep[$0.16\pm0.02 M_{\odot}$, ][]{cgk98}, J0437$-$4715
\citep[$0.236\pm0.017 M_{\odot}$, ][]{vbb+01}, J0218+4232
\citep[$0.20\pm0.02 M_{\odot}$, ][]{bvk03} and J1911$-$5958A
\citep[$0.175\pm0.010 M_{\odot}$, ][]{bvk+06} is tantalizing.


Using the Green Bank Telescope (GBT) we have searched for radio
pulsations from possible neutron-star companions to the 8 white dwarfs
listed in \citet{lbe+04} and Table \ref{tab:parms}. We describe our
observing setup in \S~\ref{sec:gbt}, our results in \S~\ref{sec:res}
and discuss our findings in \S~\ref{sec:disc}.

\begin{table}
  \centering
  \begin{tabular}{lrr}%
    \hline%
    Name     & Integration time      & Flux limit        \\
             &                   (s) &             (mJy) \\
    \hline%
    \hline%

SDSS~J002207.65$-$101423.5 &  1400 + 2800  & 0.16   \\
SDSS~J081136.34+461156.4   &  1400 + 2800  & 0.16	  \\
SDSS~J102228.02+020035.2   &  1400 + 2800  & 0.13	  \\
SDSS~J105611.03+653631.5   &  2 x 1400     & 0.21   \\
SDSS~J123410.37$-$022802.9 &  3 x 2800     & 0.19	  \\
SDSS~J130422.65+012214.2   &  3 x 1400     & 0.24	  \\
SDSS~J142601.48+010000.2   &  1400 + 2800  & 0.17	  \\
SDSS~J234536.48$-$010204.8 &  2 x 2800     & 0.15   \\

\hline%
\end{tabular}
\caption{\label{tab:parms}
Upper
  limits were derived using the longest individual integration time,
  a 22\,K receiver temperature, sky temperatures from \citet{hks+82}
  using a spectral index of $-2.6$, a minimum detection SNR of 10, a
  2.0\,K Jy$^{-1}$ telescope gain and a constant 10\% pulsar duty
  cycle.
}
\end{table}

\section{Observing setup}\label{sec:gbt} 
All sources were observed with the GBT on two or more occasions
between October 21 and 26, 2004. At a central observing frequency of
340\,MHz the Spigot backend \citep{kel+05} provided 50 MHz of
bandwidth divided into 1024 spectral channels, recording total
intensity in 16 bits
every 81.92 $\mu$s. Sessions started with 1-minute observations of
bright test pulsar PSR~J1012+5307 or PSR~J2317+1439, followed by several 23-minute
($2^{24}$ samples) or 46-minute ($2^{25}$ samples) searches in the
direction of the white dwarfs listed in Table \ref{tab:parms}.

Over the following months the data were processed using the SIGPROC
package \citep{lor01} on the 
GASP computer cluster \citep{drb+04b}. As described in more detail
below, data were examined for radio interference, corrected for
interstellar dispersion and binary acceleration, and searched for
periodic signals.

Each observation was checked for interference by summing all frequency
channels at zero dispersion measure (DM), searching for the strongest
periodicities in the resulting time series and then localizing the
matching interference within the data. Usually several affected
frequency channels or data segments had to be removed, mostly
containing non-periodic bursts or (harmonics of) 1.266 and 60\,Hz
interference. The source of the former frequency is unclear.

We next formed multiple trial time series for DMs between
5\,pc\,cm$^{-3}$ and twice the maximum expected DM in the direction of
each white dwarf, which is on average around 40\,pc\,cm$^{-3}$ \citep{cl02},
 using 0.1 to 0.5\,pc\,cm$^{-3}$ intervals.

\begin{center}
\begin{figure}
\includegraphics[]{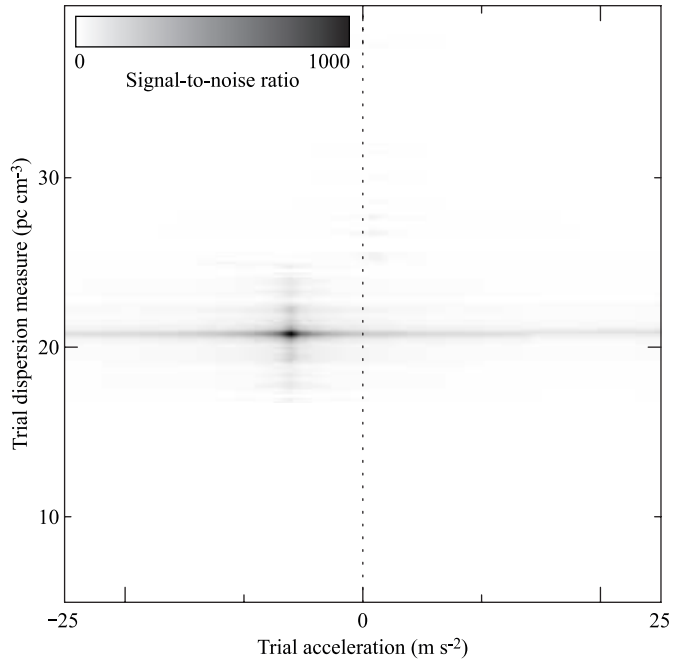}
\caption{ 
  \label{fig:1012} 
  Detection signal-to-noise versus trial dispersion measure and
  acceleration for a partly simulated 40-minute observation of
  PSR~J1012+5307. Using the PSR~J1012+5307 binary parameters from
  \citet{lcw+01}, we extrapolated one of our 1-minute test observations
  to a 40-minute observation around orbital phase 0.25. These data were
  subsequently processed in our search pipeline which detected the pulsar
  at an acceleration of 4.8\,m\,s$^{-2}$ and a DM of
  9.0\,pc\,cm$^{-3}$, close to the literature value of 9.0233 \citep{nll+95}.
 }
\end{figure}
\end{center}

We are specifically targeting millisecond pulsars in binary
systems and the acceleration that accompanies the pulsar's orbital motion
can introduce significant changes in the apparent pulse period,
hampering detection of a periodic pulse signal.  The maximum
acceleration of 90\% of all currently known pulsars falls within the
$-$25\,m\,s$^{-2}$ to $+$25\,m\,s$^{-2}$ range \citep{mhth05} and we
therefore resample each dedispersed time series for 250 trial
accelerations throughout this range.

For each dedispersed and resampled time series we then compute the
discrete Fourier transform and look for peaks, caused by periodic
signals. As different pulse shapes and duty cycles produce peaks at
different harmonics of the pulse frequency, we also fold the spectrum
2,4,8 and 16 times to add these harmonics.  For each peak in these 5
raw and folded spectra that has a signal-to-noise ratio (SNR) exceeding
5\,$\sigma$, the signal frequency and SNR are stored.

After candidates have been identified in this fashion for all trial
dedispersion and acceleration values, occurrences of the same period
at different DMs and accelerations are combined to summarize and plot
the signal properties. The first diagnostic is a two dimensional plot
of SNR versus DM and acceleration, as shown in Figure \ref{fig:1012}
for test pulsar PSR~J1012+5307. In order of peak SNR these plots are
inspected for a well-defined response in both DM and acceleration.
For the second diagnostic the time series are dedispersed,
de-accelerated and folded at the best values for DM, acceleration and
period. This allows us to check on pulse shapes, signal bandwidth
and signal transience or persistence.


\section{Results}\label{sec:res}
\begin{center}
\begin{figure}[t]
\includegraphics[]{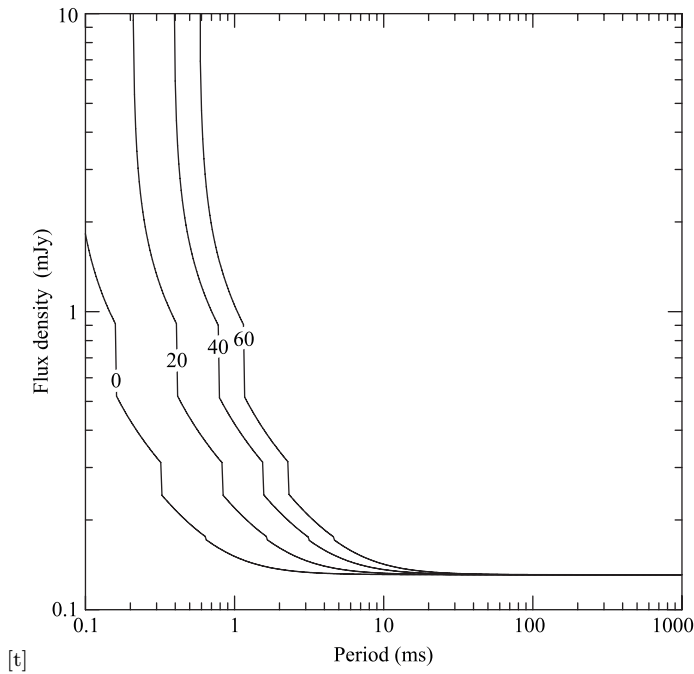}
\caption{ 
  \label{fig:sens} 
  Estimated minimum detectable flux S$_{min}$ as a function of period,
  for 4 characteristic dispersion measures (0, 20, 40 and
  60\,pc\,cm$^{-3}$). Curves are calculated for the observing setup
  described in \S~\ref{sec:gbt} and Table \ref{tab:parms}, for a
  2800\,s integration time and 30\,K sky temperature.}
\end{figure}
\end{center}

None of the observations were found to contain signals that merited
follow-up. In Table \ref{tab:parms} we estimate the pulsar-flux upper
limits for the longest observation of each candidate. The upper limits
quoted are for pulsars with periods of 20\,ms or more, for which our
search algorithm is most sensitive; in Figure \ref{fig:sens} we show
the typical dependence of this minimum detectable flux S$_{min}$ on
the pulsar period and dispersion measure.


~\\
~\\

\section{Discussion}\label{sec:disc} 
There are several evolutionary scenarios thought to produce low-mass
helium-core white dwarfs. In all these the progenitor of the observed
white dwarf interacts with a close binary companion and loses its
envelope. In one channel of evolution, the progenitor's companion is a
heavier white dwarf and a common-envelope phase exposes the
progenitor's helium core. As the companion white dwarf is older and
cooler it is much dimmer and harder to detect than the observed white
dwarf. In the other channel of evolution the progenitor of the
observed white dwarf (WD) is in a binary with a neutron star (NS). As the
progenitor evolves onto the red giant branch it overflows its Roche
lobe and a phase of mass transfer starts that both recycles the pulsar
and removes part of the progenitor's hydrogen envelope
\citep[cf.][]{bv91}.

We will outline the factors that determine radio pulsar detectability
to estimate the probability $P_{MSP}$ that a individual helium-core
white dwarf was formed in the presence of a neutron star, as opposed
to formation near another white dwarf.

We first evaluate our luminosity completeness. The combination of our
radio-flux upper limits with the white-dwarf distance estimates allows
us to constrain the radio-pulsar luminosities. Typically, our
candidate white dwarfs are 3--4 magnitudes brighter in $r$ than the
companion to PSR~J0218+4232 \citep[18.2--20.5 versus
  23.9, ][]{elh+06,bvk03} and hotter by a factor of 2.
We therefore use the low end of the distance range to PSR~J0218+4232 
\citep[2.5--4 kpc, ][]{bvk03}
to estimate the distance for each candidate to be 2 kpc. At this distance
our radio-flux upper limits amount to a luminosity of 0.6--0.8 mJy
kpc$^{-2}$. As 95\% of currently known field millisecond pulsars are
brighter than this \citep{mhth05} we assume we are complete to a level
of $P_{lum}$=0.95.

Millisecond pulsars generally have relatively wide beams that show up
as high duty cycles ($\sim$20\% on average) and increase the
chance that the beam sweeps over the Earth. For millisecond pulsars
this beaming fraction $P_{beam}$ is thought to be as high as 0.7$\pm$0.2
\citep{kxl+98}.

The maximum acceleration of 90\% of all currently known pulsars falls
in our acceleration search range, and as the 10\% with higher
maximum accelerations are still detectable during large parts of
their orbit, where the component of their motion in our line of sight
is less, we estimate $P_{acc}$=0.95.

As more than half of millisecond pulsars with known characteristic ages
appear to be older than 5 Gyr \citep{mhth05} we assume recycled radio
pulsars remain active over the cooling lifetime of white dwarfs.

We will use our non-detections and the luminosity, acceleration and
beaming fractions mentioned above to estimate the probability
$P_{MSP}$ that a low-mass helium-core white dwarf is made through the
channel involving a millisecond-pulsar companion. We assume that our
N=8 candidates were correctly identified as low-mass helium white
dwarfs.  Taking radio frequency interference
into account we estimate that our pulsar search-algorithm success rate
$P_{eff}$ is 80\%.

\vspace{2cm} 
In this case the assumption that our non-detection was the most
probable outcome of our experiment:
\begin{equation}
(1 - 
  P_{lum} \times P_{acc} \times P_{beam} \times P_{eff} \times P_{MSP} 
)^N > \frac{1}{2} 
\end{equation}
leads us to conclude that $P_{MSP} < 0.18\pm0.05$.

Our upper limit for $P_{MSP}$ is the first measurement of the
probabilities in the formation of low-mass helium-core white dwarfs.
In comparison, for general white dwarfs the mass-ratio distribution
appears to peak around unity, strongly favoring WD-WD over WD-NS
binaries: \cite{ps06} find that 50\% of detached eclipsing binaries in
the Small Magellanic Cloud have very similar masses and \cite{mmm02}
show that WD-WD binary stars relatively often have mass ratios close
to 1. Simulations by \cite{sfm+01} predict the birth rate for general
WD-WD binaries to be ~100 times higher than for WD-NS binaries -- a
significant fraction of potential WD-NS binaries are disrupted in the
supernova that creates the neutron star.

\section{Conclusions}\label{sec:con} 
We have searched for radio pulsars around 8 low-mass white
dwarf candidates from the SDSS and have found none down to a limit at
the 5th percentile of the currently known millisecond-pulsar
luminosity distribution. Low-mass helium-core white dwarfs can form
around either a heavier white dwarf or a neutron star: we estimate the
probability of formation through the latter evolution channel to be $<
0.18\pm0.05$.

\section{Acknowledgements}
Van Leeuwen and Ferdman wish to thank HIA for travel support. Stairs
holds an NSERC UFA. Pulsar research at UBC is supported by a Discovery
Grant. 
The National Radio Astronomy Observatory is a facility of the National
Science Foundation operated under cooperative agreement by Associated
Universities, Inc.


\end{document}